\documentclass[12pt,letterpaper]{amsart}
\usepackage{amssymb,latexsym, amsmath}
\usepackage[dvips]{graphics}

\theoremstyle{plain}

\theoremstyle{definition}

\theoremstyle{remark}

\numberwithin{equation}{section}
\numberwithin{theorem}{section}
\numberwithin{table}{section}
\numberwithin{figure}{section}

\newcommand{\mma}{Mathematica}

\def\({\left(}
\def\){\right)}

\begin{document}
\title{On the neighbor spacing of eigenvalues of unitary matrices}
\author{David~W.~Farmer}

\thanks{
Research supported by the
American Institute of Mathematics and the NSF focused research group grant DMS0244660}

\thispagestyle{empty}
\vspace{.5cm}
\begin{abstract}
We describe a subtle error which can appear in numerical calculations involving
the spacing statistics of eigenvalues of random unitary matrices.
\end{abstract}

\address{
{\parskip 0pt
American Institute of Mathematics\endgraf
farmer@aimath.org\endgraf
}
  }

\maketitle

\section{Introduction}

The Random Matrix Theory (RMT) of the classical compact groups
has become a central tool in analytic number theory.  By modeling
the Riemann zeta-function and other $L$-functions with the
characteristic polynomial of a random unitary matrix,
new conjectures have been made about $L$-functions, which have
also been useful for proving new results.
One of the
reasons for this success is that many quantities which are mysterious
in number theory can be computed exactly in RMT.
Examples include the statistical behavior of the zeros~\cite{Mon,Odl,KSa,Rub},
value distribution~\cite{KS1,KS2}, and moments~\cite{CFKRS}.

There are some questions about $L$-functions for which the RMT analogue has
not yet been answered.  But in many of those cases one can exploit
the fact that it is easy to generate Haar-distributed random matrices
for the unitary, symplectic, and orthogonal groups. See Mezzadri~\cite{M}
for a complete discussion.  Thus, one can usually obtain numerical
evidence to support a conjecture or to check a calculation.

In many cases, data generated from relatively small matrices is
almost indistinguishable from the limiting case of large matrices.
For example, the normalized nearest neighbor spacing of eigenvalues
of $15\times 15$ random unitary matrices differs from the limiting case
of large unitary matrices
by so little that the difference is not readily visible in a histogram.

Despite the simplicity of generating random matrices, 
it turns out there is a subtle bias in the way many people
generate and perform experiments on random matrices.
I have committed this error (not realizing it was an error) many times,
and it never made a difference.  But then I tried to check the next-to-leading
order term of a fairly involved RMT calculation~\cite{DFFHMP}.  The numerics
refused to confirm the calculation, and it turned out that the numerics were wrong.
Fortunately, the error is easily avoided once you know about it.

The next section describes a simple example of the situation in which this error
occurs, and gives some data to illustrate the problem.  In section~\ref{sec:theanswer}
the underlying cause of the problem is explained. 

\section{The neighbor spacing of eigenvalues}

The sample problem concerns the neighbor spacing of the eigenvalues of random
unitary matrices.  These spacings are completely understood: we choose this
example to illustrate what can go wrong in the numerical experiments.

Suppose $U\in U(M)$ is an $M\times M$ unitary matrix.  The characteristic
polynomial of $U$ is
\begin{equation}
\Lambda_U(x) = \prod_{m=1}^M (x-e^{i\theta_m})
\end{equation}
where $\{e^{i \theta_1},\ldots,e^{i\theta_M}\}$ are the eigenvalues of $U$
and $-\pi\le \theta_1\le \cdots\le \theta_M<\pi$ are the 
\emph{eigenangles} of~$U$.
Let $\delta_m=\frac{M}{2\pi}(\theta_{m+1}-\theta_m)$ denote the
normalized neighbor spacing of the eigenangles, where we set
$\theta_{M+1} = \theta_1 + 2\pi$ so that $\delta_M$ is the ``wrap around''
neighbor spacing.  
Collectively the $M$ numbers $\delta_1,\ldots,\delta_{M}$ are $1$ on
average, because $\delta_1+\cdots+\delta_{M} =M$ by construction. 
but as commonly implemented on a computer, the individual $\delta_j$ are
\emph{not} one on average.  That is the point of this paper.

\subsection{Generating random matrices}
The computer code and data in this paper are from \mma, but the code
should be understandable without in-depth knowledge of \mma,  and
the apparent anomalies in the data have nothing to do with \mma.

The following code defines a function {\tt randunitary[M]} which
generates a Haar-random matrix in $U(m)$.  The first line loads the
\mma\ package required to generate normally distributed random numbers.
The second line sets up the generation of the  normally distributed random numbers,
and the third command implements the algorithm described by Mezzadri~\cite{M}.

\begin{verbatim}
<<Statistics`NormalDistribution`

norm = NormalDistribution[0, 1]

randunitary[M_]:=Block[{gmatrix,q,r,d,dd},
    gmatrix=Table[Random[norm]+Random[norm]*I,{i,1,M},{j,1,M}]];
    {q,r} = QRDecomposition[gmatrix];
    d=DiagonalMatrix[Table[dd=r[[j,j]];dd/Abs[dd],{j,1,M}]];
    q.d]
\end{verbatim}

Note that this code already fixes a non-obvious error that is present in the
most straightforward way to generate a random unitary matrix.  Namely,
the matrix {\tt q} in the above code should be a Haar-random matrix in~$U(M)$,
but it isn't.  The problem is that \mma's {\tt QRDecomposition} function
introduces a bias which causes the matrix {\tt q} to have fewer eigenvalues
very close to~$1$.  Multiplying by the diagonal matrix {\tt d} fixes that
problem.  This issue arises in the {\tt QRDecomposition} routine of every
computer algebra package.  See Mezzadri~\cite{M} for details.

Note also that this code is for Mathematica~5.  Some modifications may be
needed for Mathematica~6 because the functions for generating
random numbers have been changed.

The following code extracts the eigenangles and defines the normalized neighbor
spacings.  In this example we use $14\times14$ matrices.

\begin{verbatim}

M=14;
mymatrix=randunitary[M];
eigangles= (M/(2 Pi)) Sort[Arg[Eigenvalues[mymatrix]]];
delta[j_]:=eigenangles[[j+1]] - eigenangles[[j]];
delta[M]:=eigenangles[[1]] - eigenangles[[M]] + M;
\end{verbatim}

\subsection{Problems with the eigenangle spacing}

Depending on how you use it, the above code has an error.
Here is how the error manifests itself.  For fixed $j$, the average (expected)
value of {\tt delta[j]} is \emph{not equal to~1, even in the large $M$ limit}.   
While it is true that, by
construction, the average of {\tt delta[1],...,delta[M]} is identically equal to~1,
each individual {\tt delta[j]} does not average to~1.  Table~\ref{tab:firstdata}
shows the average of {\tt delta[j]} for various $j$ and $M$, averaging
over 100,000 matrices.

\begin{table}[h!tb]
\centerline{ 
\begin{tabular}{c|c|c|c}
$M=$ & 14 & 22 & 32 \cr \hline
$\delta_1$ & 0.94345      &0.94597       &0.94506 \cr
$\delta_3$ & 0.99367      &0.99549       &0.99387 \cr
$\delta_7$ & 0.99912 &0.99836      &1.00045 \cr
$\delta_{11}$ & 0.99352     &0.99926 &0.99745 \cr
$\delta_{rand}$ & 1.00260   &0.99948    &1.00147 \cr 
\end{tabular}
}
\medskip
\caption{Averages of $\delta_j$ for random matrices in $U(M)$ for various $j$ and
$M$.  
Each column is the average for 100,000 matrices.
The bottom row averages $\delta_j$ for a randomly chosen $j$ for each matrix.
}\label{tab:firstdata}\end{table}

As the table shows, at least some of the selected $\delta_j$ differ significantly from 1.  Curiously,
most are less than~1 on average.  The last row, $\delta_{rand}$, is generated by
randomly using one of $\delta_1,\ldots,\delta_M$ for each matrix.  and then averaging
those random choices. We know that $\delta_{rand}$ must average to~1, and the closeness
of $\delta_{rand}$ to~1 should suggest that (at least some of) the other values
differ significantly from~1, as opposed to arising from fluctuations
due to a small sample size.

\subsection{More curiosities in the neighbor spacings}

The example in the previous section is unrealistic, because
nobody would use only one neighbor spacing for each matrix.
Even if that gave the right answer, it would be much more efficient
to use several, or all, of the available neighbor spacings.
However, it is quite common for people to just choose the ``lazy person's
neighbor spacings''  $\delta_1,\ldots,\delta_{M-1}$.  That is, throw
away the ``wrap around'' spacing, which avoids the need to program that
as a special case.  One could argue (incorrectly!) that since Haar measure
is rotationally invariant, all the neighbor spacings are the same, and
so no bias is introduced by omitting one value.  That argument is wrong,
as we shall see.  Here is the average of $\delta_1,\ldots,\delta_{M-1}$
for the same data set used to generate Table~\ref{tab:firstdata}.

\begin{table}[h!tb]
\centerline{\small
\begin{tabular}{r|c|c|c}
$M=$ & 14 & 22 & 32 \cr \hline
average of $\delta_1,\ldots,\delta_{M-1}$ & 0.9862      & 0.99157       & 0.99419 \cr
average of $\delta_M$ & 1.1796      & 1.1768       & 1.17988 \cr
\end{tabular}
}
\medskip
\caption{ Average of $\delta_1,\ldots,\delta_{M-1}$, and the average of $\delta_M$,
for the same data set of 100,000 matrices used to generate Table~\ref{tab:firstdata}.
}\label{tab:averagesonly}\end{table}

As Table~\ref{tab:averagesonly} shows, the average of the ``lazy person's neighbor spacings''
is less than~1.  If that is true, then the average of the
``wrap around'' spacing must be greater than~1, which Table~\ref{tab:averagesonly}
also confirms.

The fact that $\delta_M$ is larger than average is not due to an error in 
the code.  It is due to a bias caused by the way we choose the eigenangles,
which we describe in the next section.

\section{How not to choose a random gap}\label{sec:theanswer}

Suppose you want to choose a random gap between the eigenangles of a random matrix,
so that each gap is equally likely.  The following procedure is incorrect:
choose a random point on the unit circle, and then select the gap which includes
that selected point.  That method introduces a bias, because large 
gaps are more likely to be selected than small gaps.  For example, if all the
eigenvalues were in one-half of the unit circle, the biggest gap would be
selected at least half the time.  That is, the expected size of a gap surrounding any given
point will be larger than average.

That bias is implicit in the way the eigenangles have been chosen
in the \mma\ code above.  
Because we restrict the eigenangles to the interval $[-\pi, \pi)$,
the ``wrap around'' gap is chosen to include the point~$-1$.   Thus,
the expected size of $\delta_M$ should be larger than average,
as the data shows.  A calculation finds that the expected size
of a gap selected in this way is the mean of the \emph{square} of the
nearest neighbor spacing.  For the large $M$ limit of eigenangles from
$U(M)$, the square of the normalized nearest neighbor spacing is
approximately $1.180$, which agrees well with the data in Table~\ref{tab:averagesonly}.
And since $g_M$ is larger than average, the rigidity in the eigenangle
spacing forces $g_1$ and $g_{M-1}$ to be smaller than average. The
effect fades away so that $g_{M/2}$ is only slightly smaller than average,
That is, for fixed $j$, the expected value of $\delta_j$ is less than 1,
although the amount less than 1 is a rapidly decreasing function of~$j$.
The average of $\delta_1,\ldots,\delta_{M-1}$ is approximately $1-0.18/M$,
but even that can be a significant difference if one is exploring the
dependence on the size of the matrix.

Thus, even if the distribution of points on the unit circle
is rotationally invariant, there is a bias introduced by 
the simple procedure of taking the {\tt Arg[]}
of the points and dealing instead with numbers  in the interval $[-\pi, \pi)$.
In particular, the shortcut of leaving out the ``wrap around'' gap can
have unintended consequences.

\end{document}